# Hiding in plain sight: insights about health-care trends gained through open health data

A. Ravishankar Rao, PhD
Fellow, IEEE
School of Computer Science & Engg.
Fairleigh Dickinson Univ, NJ, USA
dr_ravirao@hotmail.com

Daniel Clarke
Member, IEEE
School of Computer Science & Engg.
Fairleigh Dickinson Univ, NJ, USA
danieljbclarke@gmail.com

## ABSTRACT

The open data movement constitutes an approach to achieving accountability for government organizations, and is aligned with one of the sustainable development goals outlined by the United Nations. In the area of health care, government agencies at the Federal and State levels have released open health data consisting of de-identified patient outcomes, costs and ratings. We have applied big data analytics to understand patterns and trends in open health data. We envision the use of this data by concerned citizens to understand both national and local trends in health expenditures.

We have built an open-source tool, BOAT (Big Data Open Source Analytics Tool, https://github.com/fdudatamining) to facilitate analytical exploration of open health data sets. We used BOAT to analyze data from the New York Statewide Planning and Research Cooperative System and determined that there has been a significant increase (40 percent) in the incidences of mental health issues amongst adolescents from 2009-2014. Using BOAT we analyzed costs for hip replacement surgery for 168,676 patients is in New York State, and showed that 88% of these patients had surgery costs of less than $30,000. This figure provides a basis to understand the decision by The California Public Employees' Retirement System to cap hip replacement reimbursements at $30,000, resulting in significant savings.

Our tool could enable researchers, hospitals, insurers and citizens to obtain an unbiased view on health-care expenditures, costs and emerging trends. Our tool is especially valuable in the current economic environment, where a significant amount of reporting is controlled by special interests groups and lobbies.

## 1.INTRODUCTION

The Sustainable Development Goals [1] charted by the United Nations calls for effective, accountable and inclusive institutions at all levels. The open data movement is one way of achieving accountability [2]. In this paper, we examine the arena of health due to its prominence in most national budgets. Many governments are releasing significant amounts of health data to the public, including hospital ratings, practitioners' qualifications, costs and patient outcomes [3-6]. We have applied big data analytics to understand patterns and trends in open health data. We envision the use of this data by concerned citizens to understand both national and local trends in health expenditures. Citizens could also use this data to choose hospitals and physicians and evaluate care options. Major challenges need to be overcome, including scraping and merging data from disparate sources and applying interpretive analytics [7].

We have built an open-source tool, BOAT (Big Data Open Source Analaytics Tool, https://github.com/fdudatamining) [8, 9] to facilitate analytical exploration of open health data sets. This tool could enable researchers, hospitals, insurers and citizens to obtain an unbiased view on health-care expenditures, costs and emerging trends. Our tool is especially valuable in the current economic environment, where a significant amount of reporting is controlled by special interests groups and lobbies. For instance, even scientific studies published in peer-reviewed journals have been funded by special-interest groups such as the sugar lobby [10] or the alcohol lobby [11]. It is thus important for citizens to be able to access and process the raw data, and come to their own conclusions. Our tool utilizes the Python ecosystem, and contains modules including databases, analytics, machine learning and visualization.

Though much information about medical costs and procedures is available publicly, citizens face challenges in accessing it and utilizing it effectively. Our open source tool, BOAT has yielded promising early results on open health data. We expect open-source tools to be transformative, allowing citizens to explore available data, and to arrive at their own conclusions devoid of reporting bias from other parties.

## 2. METHODS

In our earlier papers, we presented the architecture of an open-source system for the analysis of open health data[8, 12]. We use a Python-based solution with the following components: MySQL, Python Pandas, Scikit-Learn and Matplotlib [8]. Details may be found in [13]. Figure 1 depicts a high-level overview of our system.

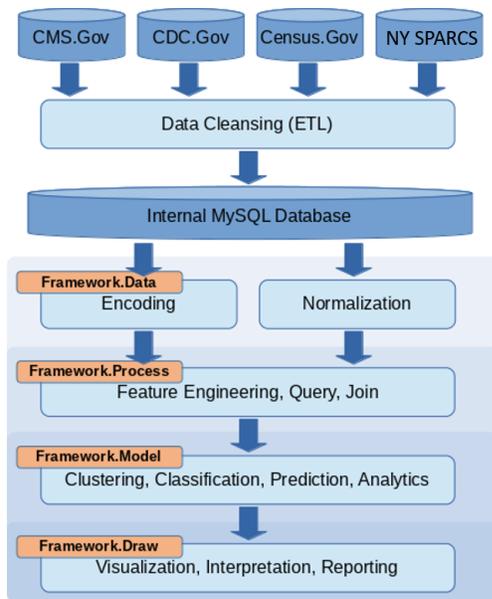

**Figure 1: A schematic illustrating the components used within our framework, BOAT, for analyzing open health data.**

We used data from the New York Statewide Planning and Research Cooperative System (SPARCS) [14] containing patient discharges and the cost of medical procedures in the state of New York. This data was subjected to operations such as grouping by labels, binning, sorting and statistical analysis to produce the results in the following section. The data is available over multiple years, allowing us to perform trend analysis.

## 3. RESULTS

We used BOAT to analyze SPARCS data [5] and examined the distribution of costs for the age group 0-17. This is shown in Figure 2. We proceeded to analyze "Mood disorders" which is one of the top three diagnoses in terms of costs. We determined that there has been a significant increase (40 percent) in the incidences of mental health issues amongst adolescents from 2009-2014. It was only in 2017 that reports from the Center for Collegiate Mental Health (http://ccmh.psu.edu/) about the increases in mental health issues on college campuses started receiving widespread attention [15]. This shows that it is important to connect the dots and understand health issues from multiple perspectives. We note that the New York SPARCS data is freely available to citizens, whereas data from other agencies including NGOs may be protected.

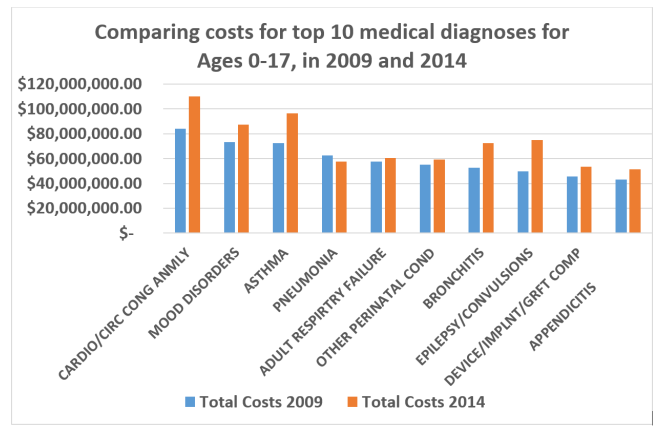

**Figure 2: Using SPARCS data from New York state to compare the costs for different medical diagnoses for the age group 0-17, from years 2009 and 2014. The top expenditure was for the diagnosis "LIVEBORN", with costs of $1.1 Billion in 2009 and $1.45 Billion in 2014. The next 10 highest total costs and their associated medical diagnoses are shown.**

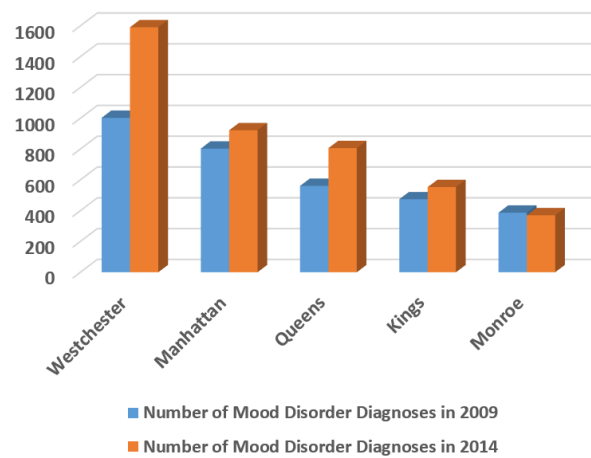

**Figure 3: Change in mood disorder diagnoses (number of cases) from 2009 to 2014. The top 5 counties in New York state in terms of cases are shown.**

The rise in the number of cases of mood disorders is shown in Figure 3. The SPARCS data show that this rise is also accompanied by a rise in the cost per case. For instance, the top hospitals in Westchester county in terms of cases of mood disorders treated were New York Presbyterian and Westchester Medical Center. New York Presbyterian experienced an increase from 2009 to 2014 of 42% in the number of cases, whereas Westchester Medical Center saw an increase of 60%. The average costs for mood disorders at these hospitals grew by 41% and 40% respectively.

We used New York SPARCS data to analyze the distribution of costs for hip replacement surgery across the state. This is shown in Figure 4. The mean cost for 168,676 patients is $22,700 and the standard deviation is $20,900. 88% of these patients had hip replacement costs of less than $30,000. The California Public Employees' Retirement System capped hip replacement reimbursements at $30,000, resulting in significant savings [16]. However, they did not reveal the reason they chose this particular figure. Our analysis through open data from a state with similar demographics shows that this cap appears reasonable.

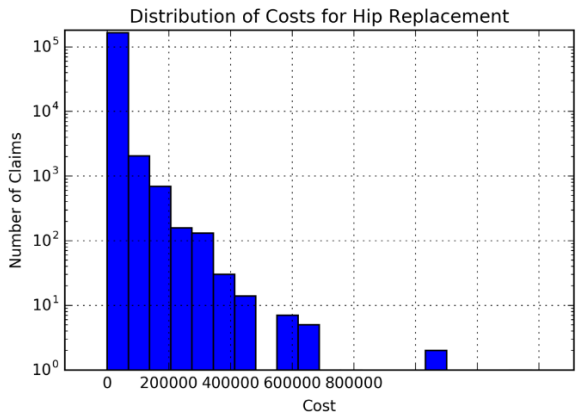

**Figure 4. The distribution of the costs for hip replacement surgery, based on New York SPARCS data for 2014.**

## 4. DISCUSSION

According to Jetzek [2], open data initiatives are yet to attain their true potential. This is echoed by the recent report [17] noting that open government data has had limited impact which does not live up to initial expectations. This raises questions regarding whether we have the right methods and intellectual tools to understand and measure the value that is contained in the data.

Based on the research presented in this paper, it appears that we are still early in the value-determination phase. We need to create the right tools, disseminate them, and have them used by the researchers and concerned citizens. Only then can we determine whether or not sufficient value exists in the open health datasets. This issue is still quite controversial and recent articles have debated both sides [18].

There have been several studies to determine the effect of releasing open health data on medical professionals and patients alike. Consider the National Health Service in the UK, which publicly releases the outcomes of cardiac surgery. A study by Bridgewater et al. [19] showed that mortality rates were reduced in regions of the UK that reported cardiac surgery outcomes. This study was not able to establish causality, though it is quite possible that a surgeon will choose to not operate on a high-risk patient, knowing that the outcome will be publicly reported.

A recent study showed that doctors' decisions are not influenced by knowing the prices of the clinical tests that they prescribe [20]. Interestingly, when prospective patients are provided price information, they appear to select options with higher rather than lower prices [21, 22]. This may be due to a perception that higher prices equate to better quality [22]. However, it has also been shown that higher spending by physicians is not associated with better patient outcomes [23].

The release of open payment data under the Federal Open Payments Program[6] [24] appears to be having the intended effect to reduce conflicts of interest between pharmaceutical companies and physicians. The Center for Medicare and Medicaid Services (CMS) in the USA is publicly releasing data about payments made to doctors. Pharmaceutical companies have responded by reducing their budget for physician meals and office visits conducted by sales representatives [24].

A central component of our research consists of an Open Source Big-Data Analytics Tool which is available freely to the research community (via github.com/fdudatamining/framework). We propose Open Source in healthcare data analytics as one of the next frontiers of innovation in healthcare [25]. Open-source shareable code and algorithms can be utilized by researchers to accelerate their investigations. The wave of the future in healthcare includes open health records [26] and open source [25].

## 5. CONCLUSION

The open source tool that we have released for big-data analytics has shown promising results when applied to open health data. We have been able to identify interesting trends in mental health diagnoses based on de-identified patient data released by New York state. We observe a significant increase of about 40% in mood disorders for the age group 0-17 years in the New York SPARCS data. This result is consistent with those obtained by other data sources, such as the Center of Collegiate Mental Health, which have more barriers for access.

Our tool also enables citizens to understand decisions made by healthcare organizations and insurance companies, especially in determining the cutoff levels for various reimbursements. We show that a cap of $30,000 for hip replacement surgery appears reasonable based on our analysis of the distribution of hip-replacement costs through New York SPARCS data.

Further work is required to accelerate the adoption of such open source tools and frameworks, and to improve the accessibility of open health data by the public.